\documentclass[twocolumn,showpacs,amsmath,amssymb,prl,superscriptaddress,floatfix]{revtex4}

\usepackage{graphicx}
\usepackage{bm}
\usepackage{amsmath,amssymb}

\begin{document}

\title{Efficient all-optical switching using slow light within a hollow fiber}

\author{M.~Bajcsy}

\author{S.~Hofferberth}
\author{V.~Balic}
\affiliation{Harvard-MIT Center for Ultracold Atoms, Department of Physics, Harvard University, Cambridge, MA 02138}
\author{T.~Peyronel}
\affiliation{Harvard-MIT Center for Ultracold Atoms, Department of Physics, MIT, Cambridge, MA 02139}
\author{M.~Hafezi}
\author{A.~S.~Zibrov}
\affiliation{Harvard-MIT Center for Ultracold Atoms, Department of Physics, Harvard University, Cambridge, MA 02138}
\author{V.~Vuletic}
\affiliation{Harvard-MIT Center for Ultracold Atoms, Department of Physics, MIT, Cambridge, MA 02139}
\author{M.D.~Lukin}
\affiliation{Harvard-MIT Center for Ultracold Atoms, Department of Physics, Harvard University, Cambridge, MA 02138}

\date{\today}

\begin{abstract}
We demonstrate a fiber-optical switch that is activated at tiny energies corresponding to few hundred optical photons per pulse. This is achieved by simultaneously confining both photons and a small laser-cooled ensemble of atoms inside the microscopic hollow core of a single-mode photonic-crystal fiber and using quantum optical techniques for generating slow light propagation and large nonlinear interaction between light beams.
\end{abstract}

\maketitle
In analogy with an electronic transistor, an all-optical switch is a device in which one light beam can fully control another. Realization of efficient all-optical switches is a long-standing goal in optical science and engineering \cite{Boyd1992}. If integrated with modern fiber-optical technologies, such devices may have important applications for optical communication and computation in telecommunication networks. Optical switches operating at a fundamental limit of one photon per switching event would further enable the realization of key protocols from  quantum information science \cite{Zeilinger2000}.

Typically, interactions of light beams in nonlinear media are very weak at low light levels. Strong interactions between few-photon pulses require a combination of large optical nonlinearity, long interaction time, low photon loss, and tight confinement of the light beams. Here, we present a fiber-optical switch that makes use of an optically dense medium containing a few hundred cold atoms trapped inside the hollow core of a photonic crystal fiber (PCF) \cite{Allan1999} within dimensions $d$ of a few micrometers. Near atomic resonance, the interaction probability $p$ between a single atom and a single photon of wavelength $\lambda$, which scales as $p\sim\lambda^2/d^2$, approaches a few percent. In this system the properties of slowly propagating photons under conditions of electromagnetically induced transparency (EIT) \cite{Harris1997,Fleischhauer2005,Hau1999} can be manipulated by pulses containing $p^{-1} \sim 100$ photons \cite{Harris1990,Imamoglu1996,Harris1998,Lukin2001}.

Simultaneous implementation of all requirements for few-photon nonlinear optics has until now only been feasible in the context of cavity quantum electrodynamics (QED) \cite{Kimble2008}, with several experiments demonstrating nonlinear optical phenomena with single intracavity photons \cite{Kimble2005, Rempe2007, Vuckovic2008}. However, these experiments remain technologically challenging and must compromise between cavity bandwidth, mirror transmission and atom-photon interaction strength. Consequently, large nonlinearities in these systems are accompanied by substantial losses at the input and output of the cavity. Our alternative cavity-free approach uses propagating fields in a PCF. Hollow-core PCF filled with molecular gas have been used for significant enhancements of efficiency in processes such as stimulated Raman scattering \cite{Russell2002} and four-wave mixing \cite{Zheltikov2003}, as well as observations of EIT \cite{Russell2006}. Recently, both room-temperature and ultra-cold atoms have successfully been loaded into PCFs  \cite{Gaeta2006, Knize2007,Ketterle2008,Luiten2007} and observations of EIT with less than a micro-Watt control field have been reported \cite{Gaeta2006, Luiten2007}.

\begin{figure}
\includegraphics[width=\columnwidth]{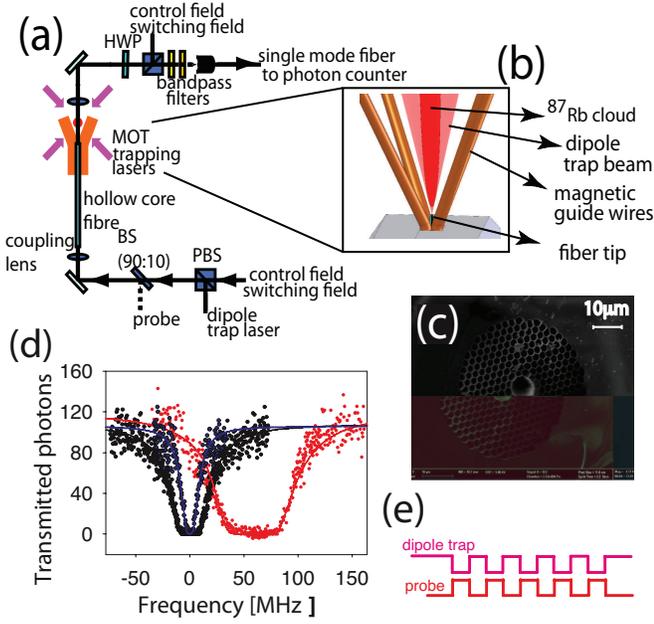}
\caption{\label{fig1}(a) Schematic of the experimental setup. (b) Schematic detail of the fiber loading. Inside the fiber, the atoms are transversely trapped by an optical trap formed by a single beam guided by the fiber. After exiting the top end of the fiber the divergent beam creates an additional funnel potential guiding the atoms into the fiber core from a distance up to $\sim100$\,$\mu$m. The atomic cloud is moved into this region while being focused with a magnetic guide consisting of four current-carrying wires that fan out from the fiber tip. (c) Scanning electron microscope image of the fiber cross section. The diameter of the central hollow region is $7$\,$\mu$m. (d) Transmission through the fiber loaded with $\sim$ 3000 atoms as a function of probe detuning from the $(5S_{1/2},F=1) \rightarrow (5P_{1/2}, F'=2)$ transition. Red points show the absorption profile with the dipole trap on continuously. Comparison to the calculated profile based on the dipole trap parameters (solid red line) confirms that the atoms are loaded inside the fiber. Black points show the observed absorption profile when probe and dipole trap are modulated as shown in (E). Here, a homogeneous broadening is caused solely by the large optical depth of the system.}
\end{figure}

Our apparatus (Fig. \ref{fig1}a) makes use of a $3$\,cm-long piece of single-mode hollow-core PCF vertically mounted inside an ultra-high vacuum chamber. A laser cooled cloud of $^{87}$Rb atoms is collected into a magneto-optical trap, focused with a magnetic guide, and loaded into the hollow core of the PCF. Once inside, the atoms are radially confined by a red-detuned dipole trap formed by a single beam guided by the fiber itself. The small diameter of the guided mode allows for strong transverse confinement (trapping frequencies $\omega_t/2 \pi \sim 50-100$\,kHz) and deep trapping potential ($\sim10$\,mK) at guiding light intensities of a few milliwatts.

To probe the atoms in the fiber, we monitor the transmission of a very low intensity ($\sim 1$\,pW) probe beam coupled into the single mode PCF (Fig. \ref{fig1}a). The signature of atoms inside the PCF are absorption profiles associated with atomic resonance lines. If the atoms are probed inside the dipole trap, we observe a unique profile shown in figure  \ref{fig1}d. The dipole trap introduces a power dependent, radially varying AC-Stark shift, which results in broadening and frequency shift of the absorption profile (red data points in Fig. \ref{fig1}d). Comparison with the calculated profile based on the dipole trap parameters verifies that the atoms are loaded inside the fiber. For the experiments described below, we avoid the broadening of the atomic transition by synchronous square-wave modulation of the dipole trap and the probe beam with opposite phase (Fig. \ref{fig1}e) at a rate much higher than the trapping frequency.
When using this technique and scanning the probe laser over a particular hyperfine transition, we observe a typical absorption profile as shown in figure \ref{fig1}d (black data points). The shape of this resonance is completely determined by the natural line profile of the transition $T_{nat}=\exp(-\mathrm{OD}/(1+4(\frac{\delta_p}{\Gamma_{e}})^2))$,
where $\Gamma_{e}$ is the lifetime of the excited atomic state and $\delta_p = \omega_{p} - \omega_{0}$ is the detuning of the probe laser from resonance. The optical depth is defined as $\mathrm{OD} = 2\,N_{eff} \,(\sigma_{eg}/A)$, where $N_{eff}$ is the effective number of atoms in the fiber (corresponding to all atoms localized on the fiber axis -- see Supplementary Information), $\sigma_{eg}$ is the atomic absorption cross section for the probed transition, and $A=\pi w_0^2$ is the beam area. Note that the optical depth for a given number of atoms inside the fiber does not depend upon the length of the atomic cloud. The measured beam waist of guided light inside the fiber is $w_0 = 1.9 \pm 0.2 $\,$\mu$m. This means that $\sim 100$ atoms inside the fiber create an optically dense medium ($OD= 1$). The profile shown in figure \ref{fig1}d yields $OD=30$, which corresponds to $\sim 3000$ atoms loaded into the fiber.

\begin{figure}
\includegraphics[width=\columnwidth]{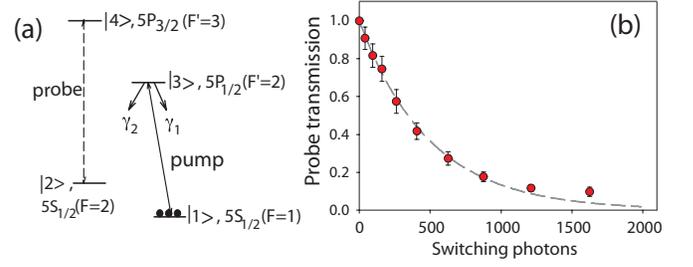}
\caption{\label{fig2}(a) Four-state atomic system for nonlinear saturation based on incoherent few-photon controlled population transfer. (b) Transmission of the probe as a function of the number of incident pump photons. $50 \%$ reduction of probe transmission is achieved with $\sim 300$ pump photons. The dashed grey line represents an exponential fit to the data.}
\end{figure}

We first consider nonlinear saturation based on incoherent population transfer in our mesoscopic atomic ensemble. The transmission of the probe beam, coupled to a cycling atomic transition, $\vert 2\rangle \rightarrow \vert 4\rangle$, is controlled via an additional pump beam transferring atoms from an auxiliary state $\vert 1\rangle$ into $\vert 2\rangle$ (Fig. \ref{fig2}a). Initially, the state $\vert 2\rangle$ is not populated, and the system is transparent for the probe beam. The incident pump beam, resonant with the $\vert 1\rangle \rightarrow \vert 3\rangle$ transition, is fully absorbed by the optically dense atom cloud, thereby transferring atoms into the $\vert 2\rangle$ state, where they then affect the propagation of the probe beam. As demonstrated in Fig. \ref{fig2}b, we achieve a $50 \%$ reduction of the probe transmission with only $300$ pump photons. The efficiency of the incoherent population transfer is limited by the branching ratio of the atomic decay from the excited state $\vert 3\rangle$. In this case, $\sim 150$ atoms are transferred into the $\vert 2\rangle$ state, which is sufficient to cause the observed absorption of the probe beam.

\begin{figure}
\includegraphics[width=\columnwidth]{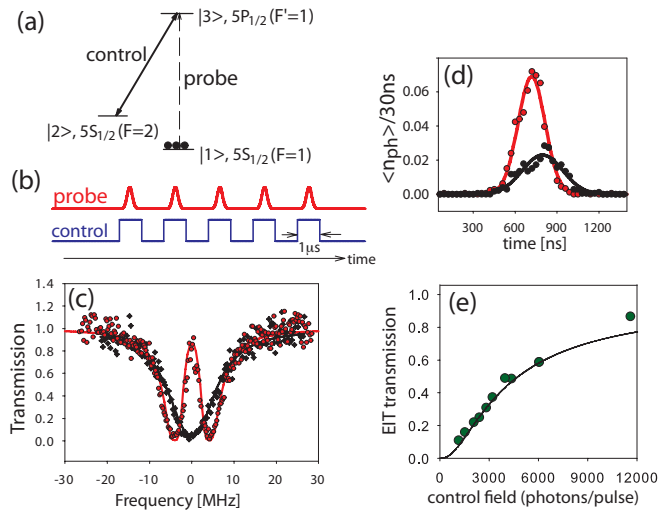}
\caption{{\label{fig3}(a) Three-level lambda scheme
used for EIT in the mesoscopic atomic medium. (b) Both probe and control field are broken into a set of $\sim100$ synchronized pulses sent through the fiber during the off-times of the dipole trap. The probe pulses have Gaussian envelopes with rms width $t_p \sim 150$\,ns. (c) Probe transmission through the fiber without (black data points) and with (red data points) control field (each data point corresponds to $100$ individual pulses. Solid lines are fits of equation (\ref{transmission}). (d) Individual probe pulse shape and delay. The EIT pulse (red) is delayed by $\sim 100$\,ns compared to the reference pulse (black), for control pulses containing $\sim 3500$ photons. Here, $\langle n_{ph}\rangle$ represents the average number of photons detected in a $30$\,ns time bin. (e) Observed transmission of the probe pulses on resonance as a function of average number of photons in the $1$\,$\mu$s control field pulse and the prediction (grey line) based on eq. (\ref{transmission})}}.
\end{figure}

Next, we demonstrate coherent interaction between few atoms and photons in our system. First, we demonstrate EIT \cite{Harris1997,Fleischhauer2005}, where atomic coherence induced by a control beam changes the transmission of a probe beam. For this we consider the three-state $\Lambda$-configuration of atomic states shown in Fig. \ref{fig3}a. In the presence of a strong control field, the weak probe field, resonant with the $\vert 1\rangle \rightarrow \vert 3\rangle$ transition, is transmitted without loss. The essence of EIT is the creation of a coupled excitation of probe photons and atomic spins ("dark-state polariton") \cite{Fleischhauer2000} that propagates through the atomic medium at greatly reduced group velocity \cite{Hau1999}.

In the absence of the control beam, our medium is completely opaque at resonance (Fig. \ref{fig3}c, black data points). In contrast, when the co-propagating control field is added, the atomic ensemble becomes transparent near the probe resonance (Fig. \ref{fig3}c, red data points). Figure \ref{fig3}d shows the profile of individual pulses and their transmission and delay due to reduced group velocity $v_g$ inside the atomic medium. For probe pulses with rms width $t_p\sim 150$\,ns we observe a group delay $t_d$ approaching $100$\,ns, corresponding to reduction of group velocity to  $v_g \approx 3$\,km/s. Remarkably, figure \ref{fig3}E shows that control pulses containing as few as $\sim 10^4$ photons are sufficient to achieve almost complete transparency of the otherwise opaque system.

\begin{figure}
\includegraphics[width=\columnwidth]{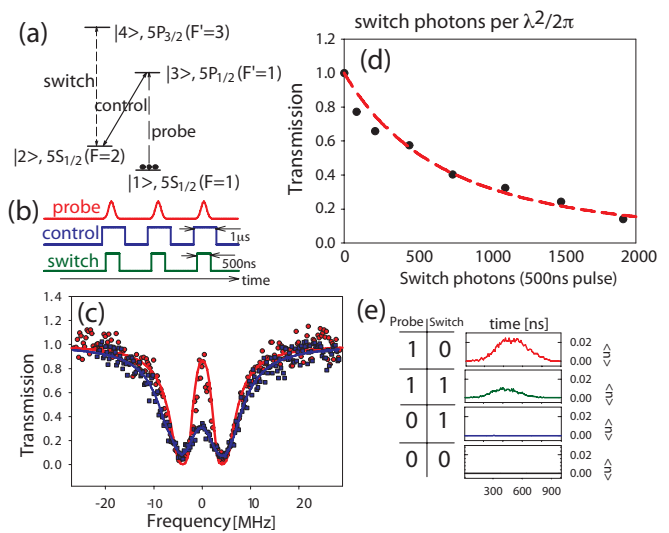}
\caption{\label{fig4}(a) Four-level system for nonlinear optical switching. The transmission of the EIT probe by a switch field resonant with an additional transition from level $\vert 2\rangle$.  (b) Timing sequences for probe, control, and switch fields. (c) Probe transmission through the fiber without (red) and with (blue) the switch field present. Solid lines are fits of equation (\ref{transmission}). (d) Observed transmission versus average number of switch photons per pulse. The solid grey line is the prediction based on equation (\ref{transmission}). The transmission is normalized to the EIT transmission in the absence of the switch photons. (e) Truth table of the switch, showing the detected photons in the output port of the switch system. Data are presented for probe pulses containing on average $\sim 2$ photons and $\sim 1/e$ attenuation of transmission in the presence of the switch photons.}
\end{figure}

The sensitive nature of the quantum interference underlying EIT enables strong nonlinear coherent interaction between the dark-state polariton and additional light fields, which can be viewed as an effective photon-photon interaction \cite{Harris1990,Imamoglu1996,Lukin2001}. An efficient nonlinear optical switch can be realized by adding to the EIT $\Lambda$-system a switch field coupling the state $\vert 2\rangle$ to an excited state $\vert 4\rangle$ (Fig. \ref{fig4}a) \cite{Harris1998}. In this scheme, the switching photons interact with flipped atomic spins (state $\vert 2\rangle$) within the slow dark-state polariton, causing a simultaneous absorption of a probe and a switch photon \cite{Zhu2001,Harris2003}. Switching is achieved when all three light fields involved (probe, control and switching field) are overlapping within the atomic cloud (Fig. \ref{fig4}b). In the absence of the switching field (red data in figure \ref{fig4}c), we observe high transmission of the probe beam on resonance due to EIT. When the switch field is turned on, this transmission is reduced. The strength of the reduction depends on the number of photons contained in the switch pulse (Fig. \ref{fig4}d). Experimentally, we observe best switching results for switch pulses of length $t_s \approx t_p + t_d$. We find a $50\%$ reduction of the initial transmission for a total number of $\sim 700$ switch photons per pulse. Figure \ref{fig4}e presents the truth table of our switch for $\sim 1/e$ attenuation of probe transmission in the presence of the switch photons. As seen from figure \ref{fig4}d, a contrast of over $90\%$ between the "on" and "off" state of the switch, desirable for practical applications, can be achieved with a moderate increase of the number of  switch photons.

We now analyze the nonlinear behavior of our atomic medium. When the resonant control and switching pulses are longer than the weak probe pulse, the effect of the atomic medium on such probe pulses with carrier frequency $\omega_p$ is given by $\mathcal{E}_{\mathrm{out}}(t) = \frac{1}{\sqrt{2\pi}}\int \mathrm{d}\omega\, \mathcal{E}_{\mathrm{in}}(\omega)\, e^{\imath \frac{\mathrm{OD}}{2}f(\omega)}\, e^{-\imath\omega t}$, where $\mathcal{E}_{\mathrm{in}}(\omega)$ is the Fourier transform of the slowly varying envelope $\mathcal{E}_{in}(t)$ of the probe pulse, normalized such that $N_{in}=\int \mathrm{d}t\,\left| \mathcal{E}_{\mathrm{in}}(t)\right|^2$, with $N_{in}$ being the number of input photons. The frequency dependent atomic response to probe light $f(\omega)$is given by \cite{Imamoglu1996,Harris1998}
\begin{equation}
\label{coherent_switch}
f(\omega) = \frac{\gamma_{13}\left(\left|\Omega_s\right|^2 - 4\delta_{12}\delta_{24}\right)}{\delta_{24}\left(4 \delta_{12}\delta_{13}- \left|\Omega_c\right|^2 \right) - \delta_{13} \left|\Omega_s\right|^2}.
\end{equation}
Here, $\Omega_{s,c}$  the Rabi frequencies of the switch and control fields The complex detunings $\delta_{ij} $ are defined as $\delta_{ij} = \delta_p + \imath \gamma_{ij}$, with $\gamma_{ij}$ being the coherence decay rates between levels $i,j$, while $\delta_p=\omega_{p} + \omega - \omega_{13}$, where $\omega_{13}$ is the frequency of the $\vert 1\rangle \rightarrow \vert 3\rangle$ transition. We consider input pulses with Gaussian envelope $\mathcal{E}_{\mathrm{in}}(t) \sim \, e^{-\frac{t^2}{4 t_p^2}}$, in which case the transmission through our atomic medium is
\begin{equation}
\label{transmission}
T(\omega_p,\mathrm{OD})= \frac{N_{out}}{ N_{in}}   =  \sqrt{\frac{2}{\pi}}\,t_p \int \mathrm{d}\omega \, e^{-2 t_p^2 \omega^2} \, e^{-\mathrm{OD} \,\mathrm{Im} f(\omega)}.
\end{equation}
We fit expression (\ref{transmission}) to observed absorption profiles such as the ones shown in figures \ref{fig3}c and \ref{fig4}c to extract the control and switch Rabi frequencies, optical depth, and ground state decoherence rate.
We next use these parameters to compare our observed EIT and nonlinear optical switch data to the theoretical prediction in figures \ref{fig3}e and \ref{fig4}d. In both cases, we find excellent agreement between our experimental data and the theoretical model. In the relevant case of resonant probe field, equation (\ref{coherent_switch}) can be approximated by
\begin{equation}
\label{switch_photons}
T = \frac{\exp\left(-N_s \left(\frac{\mu_s}{\mu_p}\right)^2 \frac{3}{\pi}\frac{ \lambda ^2}{A}\frac{t_d}{\sqrt{2} t_p+t_d} \right)}{\sqrt{1+\frac{8 t_d^2}{\mathrm{OD}\, t_p^2}}}.
\end{equation}
Here, $N_{s}$ is the number of switch photons and $\mu_{s,p}$ are switch and probe transition dipole matrix elements, respectively. Furthermore, we have assumed $\Omega_s \ll \Omega_c$, $\gamma_{12} = 0$, $\gamma_{13} \approx \gamma_{24}$ and used $t_d = L/ v_g = \frac{\mathrm{OD} \gamma_{13}}{\left|\Omega_c\right|^2}$, with $L$ being the length of the medium. For the investigated case of a relatively weak probe transition and resulting $\mathrm{OD} \approx 3$, the delay time is small ($t_d \sim t_p$) and the probe pulse is never fully stored inside the medium. If the OD is increased (either by improving the atom loading efficiency or using a stronger probe transition), $t_d \gg t_p$ and the whole probe pulse is contained inside the medium as a dark state polariton in mostly atomic form. It then follows from equation (\ref{switch_photons}) that a threshold number $N_s \sim {A \over \lambda^2}$ of switch photons causes $1/e$ attenuation of the probe pulse. This ideal limit can be easily understood. In the case of a single slow probe photon, the polariton contains only a single atomic spin at any time. Consequently, absorption of a single switch photon, which occurs with probability $p\sim {\lambda^2 \over A}$, is required to destroy this coherent atomic excitation.

Our experimental demonstrations introduce a novel physical system that opens up unique prospects in quantum and nonlinear optics. Our system can be used to implement efficient photon counting \cite{Imamoglu2002,Kwiat2002} by combining photon storage with spin-flipped atom interrogation via the cycling transition. Further improvements in nonlinear optical efficiency are possible by either simultaneously slowing down a pair of pulses to enable long interaction time \cite{Lukin2000} or using stationary-pulse techniques \cite{Lukin2003,Lukin2005}. In the former case, the threshold number of switch photons would be reduced to $N_s \sim{1 \over \sqrt{OD}} {A \over \lambda^2}$. \cite{Lukin2000}, while in the latter case the probability of interaction between two single photons scales as $\sim OD{ \lambda^2 \over A}$ \cite{Lukin2005}. Thus, with a relatively modest improvement of atom loading resulting in optical depth $OD>100$, achieving nonlinear interaction of two guided photons appears within reach. Finally, the present demonstration opens up the possibility to create strongly interacting many-body photon states \cite{Demler2008}, which may give new insights into the physics of non-equilibrium strongly correlated systems.

We thank Yiwen Chu and David Brown for their contribution to the experiment. This work was supported by Harvard-MIT CUA, NSF, DARPA and Packard Foundation.

\end{document}